\input{aipcheck}

\documentclass[
    ,final            
  ]
  {aipproc}

\layoutstyle{8x11double}

\begin{document}

\title{Constraining Palatini cosmological models using GRB data.}

\classification{98.80.-k, 04.50.Kd}

\keywords{modified gravity, cosmological simulations, dark energy theory, cosmic singularity}

\author{Micha{\l} Kamionka}{address={Astronomical Institute, University of Wroc{\l}aw\\ ul. Kopernika 11, 51-622 Wroc{\l}aw, Poland.\\ e-mail: kamionka@astro.uni.wroc.pl\\}}

\begin{abstract}
New constraints on previously  investigated Palatini cosmological models \cite{Borowiec:2011wd} have been obtained by adding Gamma Ray Burst (GRB) data \cite{Tsutsui:2012sp}.
\end{abstract}

\maketitle


\section{Cosmology from the generalized Einstein equations}

 Recently, we have investigated  cosmological applications and confronted them against astrophysical data  the following class of gravitational Lagrangians:

\footnotesize\begin{eqnarray}\label{lagr_ii2}
L=\sqrt{g}\left(f(R)+ F(R)L_d\right)+L_{mat}\equiv \cr
\equiv\sqrt{g}\left( R+\alpha R^2+\beta
R^{1+\delta}+ R^{1+\sigma}L_{d}\right) +L_{mat}
\end{eqnarray}
\normalsize
within the first-order Palatini formalism \cite{Borowiec:2011wd}.
Here $L_{d}=-\frac{1}{2}g^{\mu \nu }\partial _{\mu }\phi \partial_{\nu }\phi$ is a scalar (dilaton-like) field Lagrangian non-minimally coupled to the curvature and $L_{mat}$ represents perfect fluid Lagrangian for a dust (non-relativistic) matter.
The numerical parameters $\alpha, \beta, \delta, \sigma$ are to be determined by astrophysical data.

Applying (Palatini) variational principle compiled with flat FLRW metric one arrives to general Friedmann equation:

\footnotesize\begin{eqnarray}\label{ab11}
H^2=\frac{2(f^\prime +F^\prime L_d)\left[3f-f^\prime R+(3F-F^\prime R)L_d\right]}{3\left[2f^\prime -4F^\prime L_d +\frac{3[2f-f^\prime R +(F^\prime R-F)L_d][f^{\prime\prime}+(F^{\prime\prime}-2F^{-1}(F^\prime)^2)L_d]}
{f^{\prime\prime}R-f^\prime+[F^{\prime\prime}R +2F^\prime -2F^{-1}(F^\prime)^2R]L_d}\right]^{2}}
\end{eqnarray}
\normalsize
where $H=\frac{\dot{a}}{a}$ denotes the Hubble parameter related to the FLWR cosmic scale factor.
This reconstructs the $\Lambda$CDM model under the choice $f = R-2\Lambda$, $F = 0$, which is the limit
$\alpha = 0$, $\delta = -1$, $\beta = 2\Lambda$. Setting further $\Lambda=0$ leads to Einstein-de Sitter (decelerating) universe.

We want to recall that the generalized Friedmann equation under the form:
\begin{equation}\label{Friedmann_0}
    H^2=G(a)
\end{equation}
(which is always the case for the Palatini formalism) leads to one-dimensional particle like Newton-type dynamics 
which is fully described by the effective potential $V(a)=-\frac{1}{2}a^2 G(a)$. This relevant property allows us to compare various cosmological models on the level at the effective potential  functions and the corresponding phase-space diagrams.  Particularly, the dynamics of $\Lambda$CDM model is described by $V_{\Lambda CDM}=-\frac{1}{2}(\Lambda a^2 + \eta a^{-1})$ where $\eta$ is a density parameter for the dust matter.


As it was shown in \cite{Borowiec:2011wd} the equation (\ref{ab11}) leads to two classes of cosmological models implemented by different solutions of generalized Einstein equations.

\subsection{Model I}
Solving equations of motion by
\footnotesize\begin{eqnarray}\label{standard}
R=\rho=\eta a^{-3}, \qquad \sigma =-\delta
\end{eqnarray}
\normalsize
one obtains generalized Friedmann equation under the form
\fontsize{7}{9}\begin{eqnarray}\label{H_H0_2rozw}
\left(\frac{H}{H_0}\right)^2=\frac{2+4\Omega_{0,\alpha}(1+z)^3-2\frac{1-3\delta}{\delta}\Omega_{0,\beta}(1+z)^{3\delta}}{\left[2-2\Omega_{0,\alpha}(1+z)^3-\frac{(1-3\delta)(2-3\delta)}{\delta}\Omega_{0,\beta}(1+z)^{3\delta}\right]^2}\times\\
\times\left[2\Omega_{0,m}(1+z)^3+\Omega_{0,\alpha}\Omega_{0,m}(1+z)^6-\frac{2-3\delta}{\delta}\Omega_{0,\beta}\Omega_{0,m}(1+z)^{3(\delta+1)}\right]\nonumber
\end{eqnarray}
\normalsize
where
\footnotesize\begin{eqnarray}\label{dimensionless1}
\Omega_{0,m}=\frac{\eta}{3H_0^2}, \qquad \Omega_{0,\beta}=\beta\eta^\delta, \qquad \Omega_{0,\alpha}=\alpha\eta
\end{eqnarray}
\normalsize
are dimensionless (density like) parameters.

\subsection{Model II}

Another cosmological model can be determined by
\footnotesize\begin{eqnarray}\label{news}
R=\left[\frac{\eta}{(1-\delta)\beta}\right]^{\frac{1}{1+\delta}} a^{-\frac{3 }{1+\delta}}, \qquad \sigma=2\delta
\end{eqnarray}
\normalsize

which leads to
\fontsize{7}{9}\begin{eqnarray}\label{H_H0}
\left(\frac{H}{H_0}\right)^2\!=\!\frac{\frac{1+4\delta}{\delta}+12\Omega_{0,\alpha}(1+z)^{\frac{3}{1+\delta}}
+2\frac{1+\delta}{1-\delta}\Omega_{0,m}\Omega_{0,\beta}^{-1}(1+z)^{\frac{3\delta}{1+\delta}}}
{\left[\frac{1+4\delta}{\delta}+6\frac{2\delta-1}{1+\delta}\Omega_{0,\alpha}(1+z)^{\frac{3}{1+\delta}}
+\frac{2-\delta}{1-\delta}\Omega_{0,m}\Omega_{0,\beta}^{-1}(1+z)^{\frac{3\delta}{1+\delta}}\right]^2}\\
\times\left[\frac{1+\delta}{\delta}\Omega_{0,\beta}(1+z)^{\frac{3}{1+\delta}}
+3\Omega_{0,\alpha}\Omega_{0,\beta}(1+z)^{\frac{6}{1+\delta}}+\frac{2-\delta}{1-\delta}\Omega_{0,m}(1+z)^3\right]\nonumber
\end{eqnarray}
\normalsize
where now
\footnotesize\begin{eqnarray}\label{dimensionless2}
\Omega_{0,m}=\frac{\eta}{3H_0^2},\ \Omega_{0,\beta}=\frac{1}{3H_0^2}\left[\frac{\eta}{(1-\delta)\beta}\right]^{\frac{1}{1+\delta}}, \ \Omega_{0,\alpha}=\alpha H_0^2\Omega_{0,\beta}
\end{eqnarray}
\normalsize

Both models have $\Omega_{0,m}, \Omega_{0,\alpha}, \Omega_{0,\beta}, \delta$ as free parameters. By the normalization condition $H(0)=H_0$, only three of them are independent ($H_0$ denotes the Hubble constant).

\section{Fitting parameters of the models}

In order to estimate the parameters of our models we use a sample of $N=557$ supernovae (SNIa) data \cite{union2}, the observational $H(z)$ data \cite{Simon2004}, the measurements of the baryon acoustic oscillations (BAO) from the SDSS luminous red galaxies \cite{bao}, information from CMB \cite{Komatsu:2010fb} and, as an adition to \cite{Borowiec:2011wd}, information coming from observations of GRB \cite{Tsutsui:2012sp}.

The entire likelihood function $L_{TOT}$ is characterized by:
\begin{equation}
L_{TOT}=L_{SN}L_{H_z}L_{BAO}L_{CMB}L_{GRB}.
\label{total_chi2}
\end{equation}

We have assumed flat prior probabilities for all model's parameters. We also assumed that $H_0=74.2 \ [kms^{-1}Mpc^{-1}]$  \cite{0004-637X-699-1-539}.

The likelihood function is defined in the following way:
\footnotesize\begin{equation}
L_{SN} \propto \exp \left[ - \sum_{i}\frac{(\mu_{i}^{\mathrm{theor}}-\mu_{i}^{\mathrm{obs}})^{2}}
{2 \sigma_{i}^{2}} \right]  ,\label{chi2}
\end{equation}
\normalsize
where: $\sigma_{i}$ is the total measurement error, $\mu_{i}^{obs}=m_{i}-M$ is the measured value ($m_{i}$--apparent magnitude, $M$--absolute magnitude of SNIa), $\mu_{i}^{theor}=5\log_{10}D_{Li} + \mathcal{M}=5\log_{10}d_{Li} + 25$, $\mathcal{M}=-5\log_{10}H_{0}+25$ and $D_{Li}=H_{0}d_{Li}$, where $d_{Li}$ is the luminosity distance given by $d_{Li}=(1+z_{i})c\int_{0}^{z_{i}} \frac{dz'}{H(z')}$ (with the assumption $k=0$). In this paper the likelihood as a function independent of $H_0$ has been used (which is obtained after analytical marginalization of formula (\ref{chi2}) over $H_0$).

 For the $H(z)$ data the likelihood function is given by:
\footnotesize\begin{equation}
L_{H_z} \propto \exp \left[ - \sum_i\frac{\left(H(z_i)-H_i\right)^2}{2 \sigma_i^2} \right ],
\label{hz}
\end{equation}
\normalsize
where $H(z_i)$ is the Hubble function, $H_i$ denotes observational data.

For BAO A parameter data the likelihood function is characterized by:
\footnotesize\begin{equation}
L_{BAO} \propto \exp \left[  -\frac{(A^{theor}-A^{obs})^2}{2\sigma_A^2} \right] ,
\end{equation}
\normalsize
where \footnotesize$A^{theor}=\sqrt{\Omega_{m,0}} \left (\frac{H(z_A)}{H_{0}} \right ) ^{-\frac{1}{3}} \left [ \frac{1}{z_{A}} \int_{0}^{z_{A}}\frac {H_0}{H(z)} dz\right]^{\frac{2}{3}}$\normalsize and $A^{obs}=0.469 \pm 0.017$ for $z_{A}=0.35$.

We also use constraints coming from CMB temperature power spectrum, ie. CMB $R$ shift parameter \cite{Bond:1997}, which is related to the angular diameter distance ($D_A (z_*)$) to the last scattering surface:
\footnotesize\begin{equation}
R=\frac{\sqrt{\Omega_m H_0}}{c}(1+z_*)D_A(z_*).
\end{equation}
\normalsize
The likelihood function has the following form:
\footnotesize\begin{equation}
L_{CMB} \propto \exp \left[ -\frac{1}{2}\frac{(R-R_{obs})^2}{\sigma_{A}^2} \right],
\end{equation}
\normalsize
where $R_{obs}=1.725$ and $\sigma_{A} ^{-2}=6825.27$ for $z_*=1091.3$ \cite{Komatsu:2010fb}.

The likelihood function for GRB data is defined as:
\footnotesize\begin{eqnarray}
L_{GRB} \propto \exp \left[ - \sum_{i} \left[\frac{\mu_i - \mu^{\rm th}(z_i, \Omega_m, \Omega_{\Lambda},)}
       {\sigma_{\mu_i}}\right]^{2} \right]
\end{eqnarray}
\normalsize

\begin{figure}
   \includegraphics[height=.3\textheight]{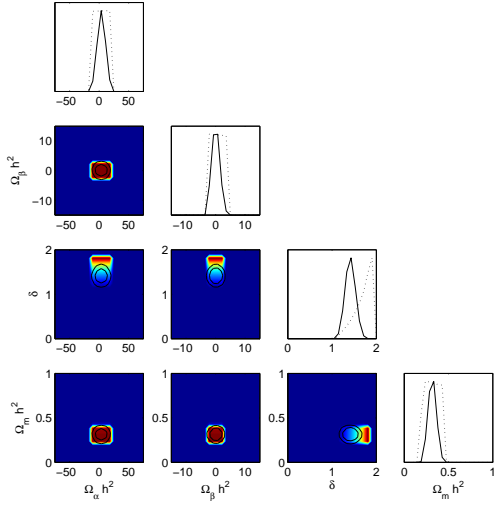}\\
   \caption{Constraints of the parameters of model $I$. In 2D plots solid lines are the 68\% and 95\% confidence intervals from the marginalized probabilities. The colors describe the mean likelihood of the sample. In 1D plots solid lines denote marginalized probabilities of the sample, dotted lines are mean likelihood. For numerical results see Table \ref{zestawienie}.}
   \label{nowy_wzor_grb_tri}
\end{figure}

\begin{figure}
   \includegraphics[height=.3\textheight]{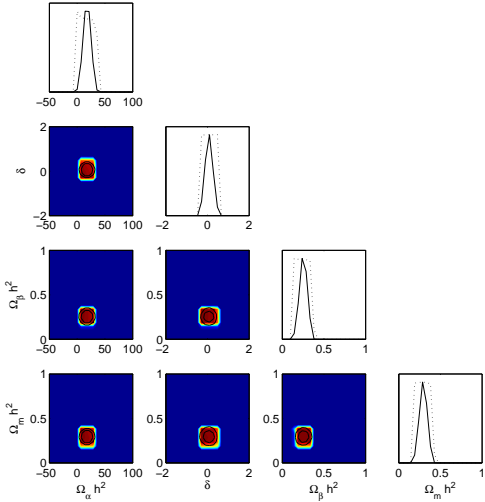}\\
   \caption{Constraints of the parameters of model $II$.  The meaning of the colors and the lines this same as in the picture \ref{nowy_wzor_grb_tri}. For numerical results see Table \ref{zestawienie}.}
   \label{nowe_dane_grb_tri}
\end{figure}

The mode of joined posterior pdf as well as mean (together with $68\%$ credible interval) of marginalized posterior pdf were calculated, by means of Markov Chains Monte Carlo analysis, using free accessible CosmoNest code \cite{Lewis1} which has been modified for our purpose. The results are presented on fig. \ref{nowy_wzor_grb_tri},\ref{nowe_dane_grb_tri}.

The numerical values of best fitted parameters for two our models as well as for $\Lambda$CDM are collected in table~\ref{zestawienie}: the previous estimations without the GRB data (i.e. SNIa, H(z) and BAO and CMB) are shown in top part of the table. The new estimations including the GRB data occupy bottom part of the table.

\begin{figure}[h]
  \includegraphics[height=.3\textheight, angle=-90]{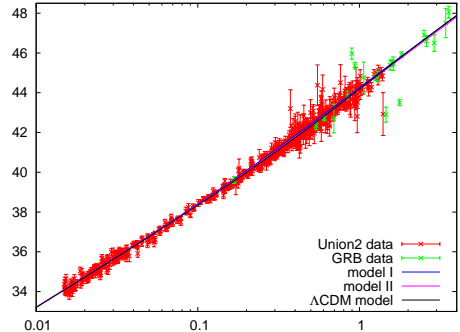}
  \caption{Comparison of Hubble's diagrams for models: I (blue), II (magenta) and $\Lambda$CDM (black).}
  \label{hubble}
\end{figure}

Quality of the estimation can be visualized on the Hubble's diagram (fig. \ref{hubble}). Both of our models are in good agreement in the observational data.

\begin{figure}
   \includegraphics[width=0.45\textwidth]{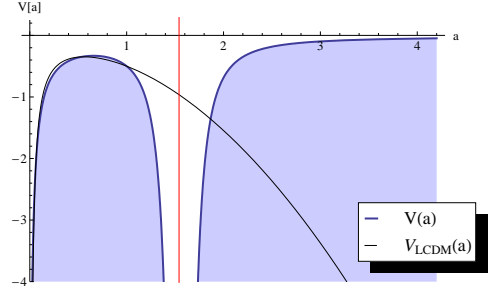}\\
   \caption{The diagram of the effective potential in particle--like representation of cosmic dynamics for model I versus $\Lambda$CDM model. Note that till the present epoch two potential plots almost coincide. Particulary, one can observe decelerating BB era. Maximum of the potential function corresponds to Einstein's unstable static solution. Discrepancies become important in the future time: e.g. discontinuities of the potential functions (vertical, red line) denote that $V \rightarrow -\infty$, i.e. $\dot a \rightarrow \infty$ for $a \rightarrow a^{final}$. It turns out to be finite--time (sudden) singularity. In any case the shadowed region below the graph is forbidden for the motion.}
   \label{model_i_potencjal}
\end{figure}

\begin{figure}
   \includegraphics[width=0.45\textwidth]{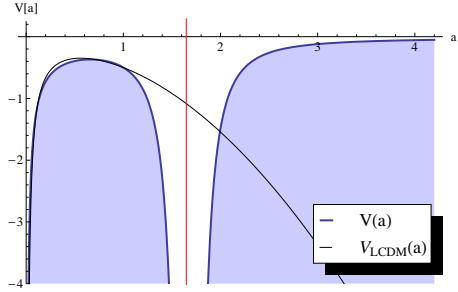}\\
   \caption{The diagram of the effective potential in particle like representation of cosmic dynamics for the model $II_{\alpha=0}$ versus $\Lambda$CDM model.  Maximum of the potential function corresponds to unstable static solution. Again, until the present epoch there is no striking differences between plots. One can observe finite--size sudden singularity in the near future (vertical, red line). In any case the shadowed region below the potential is forbidden for the motion.}
   \label{model_ii_potencjal}
\end{figure}

\section{Conclusions}

In this paper we continued and completed analysis of new cosmological models which were previously described and investigated in our paper \cite{Borowiec:2011wd}. Adding GRB data \cite{Tsutsui:2012sp} allowed us to obtain better constraints of parameter $\Omega_\alpha$ which wasn't present previously.

As it can be seen on the potential plots (fig. \ref{model_i_potencjal},\ref{model_ii_potencjal}, both models dynamically mimics $\Lambda$CDM model from the Big Bang singularity until the present time. Discrepancies will appear in the near future. Both of our models predict the final finite size and finite time singularities (at $a=1.673$ for the model I, and at $a=1.559$ for the model II). However, comparing with our previous simulations, adding new GRB data has changed properties of the model II (Big Bounce is now replaced by Big Bang).

\begin{table}[!t]
\makeatother
\tiny\begin{tabular}{cccccp{.3\textwidth}} 
\hline
   \tablehead{5}{c}{b}{models I - the parameters estimated without GRB data}\\
\hline
  \tablehead{1}{c}{b}{$\Omega_{0,\alpha}$} &
  \tablehead{1}{c}{b}{$\Omega_{0,\beta}$} &
  \tablehead{1}{c}{b}{$\delta$} &
  \tablehead{2}{c}{b}{$\Omega_{0,m}$}\\
\hline
  $-18.031^{+3.911}_{-11.969}(-6.210)$ & $5.678^{+4.322}_{-1.489}(2.190)$ & $0.238^{+0.075}_{-0.010}(0.229)$ &\multicolumn{2}{c}{$0.25\pm0.03(0.23)$}\\
\hline
   \tablehead{5}{c}{b}{models II - the parameters estimated without GRB data}\\
\hline
  \tablehead{1}{c}{b}{$\Omega_{0,\alpha}$} &
  \tablehead{1}{c}{b}{$\Omega_{0,c}$} &
  \tablehead{1}{c}{b}{$\delta$} &
  \tablehead{1}{c}{b}{$\Omega_{0,\beta}$} &
  \tablehead{1}{c}{b}{$\Omega_{0,m}$} \\
\hline
  $-44.686^{+5.016}_{-15.314}(-57.870)$ & $0.715^{+0.196}_{-0.393}(0.232)$ & $0.598^{+0.008}_{-0.011}(0.560)$ & $0.009\pm0.005 (0.003)$ & $0.05^{+0.05}_{-0.04}(0.001)$\\
\hline
  \tablehead{5}{c}{b}{model $\Lambda$CDM - the parameter estimated without GRB data}\\
\hline
  \multicolumn{5}{c}{$0.262^{+0.011}_{-0.012}(0.262)$}\\
\hline
\hline
   \tablehead{5}{c}{b}{models I - the parameters estimated using GRB data}\\
\hline
  \tablehead{1}{c}{b}{$\Omega_{0,\alpha}$} &
  \tablehead{1}{c}{b}{$\Omega_{0,\beta}$} &
  \tablehead{1}{c}{b}{$\delta$} &
  \tablehead{2}{c}{b}{$\Omega_{0,m}$}\\
\hline
  $3.809^{+0.133}_{-0.150}(3.776)$ & $0.0002^{+0.023}_{-0.0002}(0.011)$ & $1.605^{+0.083}_{-0.275}(1.413)$ &\multicolumn{2}{c}{$0.315^{+0.013}_{-0.014}(0.316)$}\\
\hline
   \tablehead{5}{c}{b}{models II - the parameters estimated using GRB data}\\
\hline
  \tablehead{1}{c}{b}{$\Omega_{0,\alpha}$} &
  \tablehead{1}{c}{b}{$\Omega_{0,c}$} &
  \tablehead{1}{c}{b}{$\delta$} &
  \tablehead{1}{c}{b}{$\Omega_{0,\beta}$} &
  \tablehead{1}{c}{b}{$\Omega_{0,m}$} \\
\hline
  $17.754^{+1.041}_{-0.840}(17.829)$ & $1.157^{+0.044}_{-0.020}(1.163)$ & $0.069^{+0.015}_{-0.012}(0.071)$ & $0.254^{+0.008}_{-0.011}(0.253)$ & $0.293^{+0.015}_{-0.012}(0.295)$\\
\hline
  \tablehead{5}{c}{b}{model $\Lambda$CDM - the parameter estimated using GRB data}\\
\hline
  \multicolumn{5}{c}{$0.260^{+0.004}_{-0.003}(0.260)$}\\
\hline
\end{tabular}
\caption{The values of estimated parameters (mean of the marginalized posterior probabilities and $68 \%$ credible intervals or sample square roots of variance, together with mode of the joined posterior probabilities, shown in brackets) for two investigated models. We compare estimations without GRB data (top part of the table) with the one employing GRB data (bottom part).}
\label{zestawienie}
\end{table}

\begin{theacknowledgments}
  M.K. is supported by the Polish NCN grant PRELUDIUM 2012/05/N/ST9/03857.
\end{theacknowledgments}


\begin{thebibliography}{9}
\expandafter\ifx\csname natexlab\endcsname\relax\def\natexlab#1{#1}\fi
\providecommand{\enquote}[1]{``#1''}
\expandafter\ifx\csname url\endcsname\relax
  \def\url#1{\texttt{#1}}\fi
\expandafter\ifx\csname urlprefix\endcsname\relax\def\urlprefix{URL }\fi
\providecommand{\eprint}[2][]{\url{#2}}

\bibitem[Borowiec et al. (2011)]{Borowiec:2011wd}
  A.~Borowiec, M.~Kamionka, A.~Kurek and M.~Szydlowski,
  ``Cosmic acceleration from modified gravity with Palatini formalism,''
  JCAP {\bf 1202}, 027 (2012), \eprint{arXiv:1109.3420}.

\bibitem[Tsutsui et al. (2012)]{Tsutsui:2012sp}
  R.~Tsutsui, T.~Nakamura, D.~Yonetoku, K.~Takahashi and Y.~Morihara,
  ``Gamma-Ray Bursts are precise distance indicators similar to Type Ia Supernovae?'',(2012), \eprint{arXiv:1205.2954}.

\bibitem[Amanullah et al. (2010)]{union2}
 R.~Amanullah {\it et al.},
  ``Spectra and Light Curves of Six Type Ia Supernovae at 0.511 < z < 1.12 and the Union2 Compilation'',
  Astrophys.\ J.\  {\bf 716}, 712 (2010), \eprint{arXiv:1004.1711}.

\bibitem[Simon et al. (2004)]{Simon2004}
 J.~Simon, L.~Verde, R.~Jimenez,
  ``Constraints on the redshift dependence of the dark energy potential'',
  Phys.\ Rev.\  {\bf D71}, 123001 (2005), \eprint{astro-ph/0412269}.

\bibitem[Eisenstein et al. (2005)]{bao}
 D.~J.~Eisenstein {\it et al.},
  ``Detection of the baryon acoustic peak in the large-scale correlation function of SDSS luminous red galaxies'',
  Astrophys.\ J.\  {\bf 633}, 560-574 (2005), \eprint{astro-ph/0501171};\\
 W.~J.~Percival, S.~Cole, D.~J.~Eisenstein, R.~C.~Nichol, J.~A.~Peacock, A.~C.~Pope, A.~S.~Szalay,
  ``Measuring the Baryon Acoustic Oscillation scale using the SDSS and 2dFGRS'',
  Mon.\ Not.\ Roy.\ Astron.\ Soc.\  {\bf 381}, 1053-1066 (2007), \eprint{arXiv:0705.3323};\\
 B.~A.~Reid {\it et al.},
  ``Baryon Acoustic Oscillations in the Sloan Digital Sky Survey Data Release 7 Galaxy Sample'',
  Mon.\ Not.\ Roy.\ Astron.\ Soc.\  {\bf 401}, 2148-2168 (2010), \eprint{arXiv:0907.1660}.

\bibitem[Komatsu et al. (2010)]{Komatsu:2010fb}
 E.~Komatsu {\it et al.},
  ``Seven-Year Wilkinson Microwave Anisotropy Probe (WMAP) Observations: Cosmological Interpretation'',
  Astrophys.\ J.\ Suppl.\  {\bf 192}, 18 (2011), \eprint{arXiv:1001.4538}.

\bibitem[Bond et al. (1997)]{Bond:1997}
 J.~R.~Bond, G.~Efstathiou, M.~Tegmark,
  ``Forecasting cosmic parameter errors from microwave background anisotropy experiments'',
  Mon.\ Not.\ Roy.\ Astron.\ Soc.\  {\bf 291}, L33-L41 (1997), \eprint{astro-ph/9702100}.

\bibitem[Riess et al. (2009)]{0004-637X-699-1-539}
 A.~G.~Riess {\it et al.},
  ``A Redetermination of the Hubble Constant with the Hubble Space Telescope from a Differential Distance Ladder'',
  Astrophys.\ J.\  {\bf 699}, 539 (2009), \eprint{arXiv:0905.0695}.

\bibitem[Mukherjee et al. (2006)]{Lewis1}
 P.~Mukherjee, D.~Parkinson, A.~R.~Liddle,
  ``A nested sampling algorithm for cosmological model selection'',
  Astrophys.\ J.\  {\bf 638}, L51-L54 (2006), \eprint{astro-ph/0508461};\\
 P.~Mukherjee, D.~Parkinson, P.~S.~Corasaniti, A.~R.~Liddle, M.~Kunz,
  ``Model selection as a science driver for dark energy surveys'',
  Mon.\ Not.\ Roy.\ Astron.\ Soc.\  {\bf 369}, 1725-1734 (2006), \eprint{astro-ph/0512484};\\
 D.~Parkinson, P.~Mukherjee, A.R.~Liddle,
  ``A Bayesian model selection analysis of WMAP3'',
  Phys.\ Rev.\  {\bf D73}, 123523 (2006), \eprint{astro-ph/0605003};\\
 \url{http://cosmonest.org/}


\end{thebibliography}
\end{document}